\documentclass[sigchi]{acmart}

\usepackage{booktabs} 

\copyrightyear{2019}
\acmYear{2019}
\setcopyright{acmlicensed}
\acmConference[CHI 2019]{CHI Conference on Human Factors in Computing Systems Proceedings}{May 4--9, 2019}{Glasgow, Scotland Uk}
\acmBooktitle{CHI Conference on Human Factors in Computing Systems Proceedings (CHI 2019), May 4--9, 2019, Glasgow, Scotland Uk}
\acmPrice{15.00}
\acmDOI{10.1145/3290605.3300336}
\acmISBN{978-1-4503-5970-2/19/05}

\usepackage{graphics}      
\usepackage{color, colortbl}
\definecolor{Gray}{gray}{0.9}
\usepackage{ccicons}          
\usepackage[utf8]{inputenc} 
 \usepackage{paralist} 
\newcommand{\comm}[1]{\textcolor{black}{#1}}
\usepackage{multirow}

\begin{document}
\title{`I make up a silly name': Understanding Children's Perception of Privacy Risks Online}

\author{Jun Zhao}
\email{jun.zhao@cs.ox.ac.uk}
\affiliation{%
  \institution{Department of Computer Science. University of Oxford}
  \city{Oxford}
  \country{UK}
}

\author{Ge Wang}
\email{ge.wang.17@ucl.ac.uk}
\affiliation{%
  \institution{Department of Information Studies. University College London}
  \city{London}
  \country{UK}
}

\author{Carys Dally}
\email{carysdally@gmail.com}
\affiliation{%
  \institution{Department of Experimental Psychology. University of Oxford}
  \city{Oxford}
  \country{UK}
}

\author{Petr Slovak}
\email{p.slovak@ucl.ac.uk}
\additionalaffiliation{
 \institution{UCL Interaction Centre, University College London }
 \city{London}
 \state{}
 \country{United Kingdom}
}
\affiliation{
 \institution{Department of Informatics,\\ King’s College London }
 \city{London}
 \state{}
 \country{UK}
}

\author{Julian Edbrooke-Childs}
\email{Julian.Childs@annafreud.org}
\additionalaffiliation{%
  \institution{Anna Freud National Centre for Children \& Families}
  \city{London}
  \country{UK}
}
\affiliation{
 \institution{UCL Interaction Centre, University College London }
 \city{London}
 \state{}
 \country{United Kingdom}
}

\author{Max Van Kleek}
\email{max.van.kleek@cs.ox.ac.uk}
\author{Nigel Shadbolt}
\email{nigel.shadbolt@cs.ox.ac.uk}
\affiliation{
  \institution{Department of Computer Science. University of Oxford}
  \city{Oxford}
  \country{UK}
}

\begin{abstract}
Children under 11 are often regarded as too young to comprehend the implications of online privacy. Perhaps as a result, little research has focused on younger kids' risk recognition and coping. Such knowledge is, however, critical for designing efficient safeguarding mechanisms for this age group. Through 12 focus group studies with 29 children aged 6-10 from UK schools, we examined how children described privacy risks related to their use of tablet computers and what information was used by them to identify threats. We found that children could identify and articulate certain privacy risks well, such as information oversharing or revealing real identities online; however, they had less awareness with respect to other risks, such as online tracking or game promotions. Our findings offer promising directions for supporting children's awareness of cyber risks and the ability to protect themselves online.
\end{abstract}

%
%



\maketitle

\section{Introduction}

Today, children are spending more time online than with other media sources, such as watching television or playing offline video games~\cite{ofcom2017,livingstone2017children}. Among the many kinds of devices now connected to the Internet, mobile devices (such as tablet computers or smartphones) have become the primary means by which children go online~\cite{livingstone2017children}. In the UK, 44\% of children aged five to ten have been provided with their own tablets, with this percentage increasing annually~\cite{ofcom2017}, while in the US, ownership of tablets by children in this age group grew fivefold between 2011 and 2013~\cite{commonsense2013}. Children under five are also using smartphone and tablets more often, as the category of apps designed for younger kids continues to expand rapidly~\cite{ofcom2017, lifewire}. 

Whilst online content has opened up significant new opportunities and experiences for children to learn and have fun~\cite{judge2015using,papadakis2016comparing,neumann2014touch,schacter2016improving}, parents and educators alike are raising concerns about the rapid adoption of online content and apps by children, in particular as it relates to the amount of time spent online, and online safety~\cite{hiniker2016screen,livingstone2018parentsc,digital2017}. A recent survey has shown that parents consider privacy their primary concern regarding their children's use of the Internet~\cite{livingstone2018parentsc}. This concern is driven, in part, by the view that children themselves are too young to understand the privacy risks they might face~\cite{livingstone2018parentsb,livingstone2018children}, which drives many parents to take a restrictive approach to filter or monitor their children's online activity~\cite{kumar2018cscw,zhang2016nosy}. Indeed, recent studies show how children's ability to recognise risks online  remains inadequate, and as a result, the children are dependent on parents' strategies to keep them safe~\cite{kumar2018cscw,zhang2016nosy,livingstone2017children, livingstone2018childrenc}.  Unfortunately, however, parents themselves often feel they lack an adequate understanding of the landscape of potential risks online---ranging from bullying to cybercrime---making them uncertain about the effectiveness of their mediation approaches~\cite{livingstone2018parentsb}. 

We feel that it is important to understand children's privacy behaviours and conceptualisations not only because it might identify conceptual gaps in understanding that lead to better ways to educate and protect children online, but also because such behaviours have been shown to be indicative of later behaviours and attitudes~\cite{digital2017, wisniewski2017parents,kumar2018cscw}.  To this end, we build on recent work examining children's perceptions of online privacy risks~\cite{kumar2018cscw,zhang2016nosy,mcnally2018co,ey2011exploring}, and extend existing literature in two ways: first, by examining in detail how children \emph{describe} certain common kinds of risks, for the purpose of understanding their conceptualisations of them; second, we examine the risk coping strategies taken by the children for each distinct risk context. Specifically, we focus on the following three research questions:

\begin{itemize}
    \item \emph{R1: Perception} - Do children care about privacy online? When they do, how do they describe privacy risks?
    \item \emph{R2: Recognition} - How do children recognise risks? What information do they use in the process? 
    \item \emph{R3: Response} - How do children apply their existing knowledge in responding to risks in different scenarios, including threats they never experienced before? What do such responses suggest about information and knowledge that might be needed?
\end{itemize}

We report our results based on 12 focus groups with 29 children, aged 6-10, from UK schools undertaken between June and August 2018. We found that children in our study had a good understanding of risks related to \textit{inappropriate content, the approach of strangers}, and \textit{oversharing of personal information online}. However, they struggled to fully understand and describe risks related to \textit{online game/video promotions and personal data tracking}. Moreover, children's risk coping strategies depended on their understanding of the risks and their previous experiences: effective risk strategies were applied only if children recognised certain risks or when they \textit{felt} something untoward. These findings demonstrate the importance of learning about potential risks through a multitude of channels, such as school, parents, friends, siblings, and personal experiences~\cite{zhang2017cyberheroes,hashish2014involving}.

\comm{We identified  significant gaps in children's current awareness and understanding of two major forms of online risk: (i) online recommendations, and (ii) implications of data tracking, i.e., the recording and sharing of data pertaining to what they do online. Children remain poorly equipped to identify targeted promotional material online, including adverts and in-app promotions, exploiting tracked activity data.} We offer recommendations on the development of tools to facilitate children's understanding of privacy risks, and the need to extend children's awareness of implicit personal data access, which is becoming ever more prevalent in the games and content they encounter~\cite{federal2015ftc,reyes2017our,Binns2018MobileEcosystem}.

\section{Background}

\subsection{Children's Perception of Privacy}

Privacy has been described as a ``concept in disarray''\cite{solove2008understanding} reflecting the complex and often overlapping conceptualisations. O'Hara proposed ``seven veils'' of privacy, spanning the conceptual, factual, phenomenological, preferential, societal, legal, and moral~\cite{ohara2016seven}. Practical research in privacy has typically spanned multiple such veils; Livingstone suggests that privacy is related to the keeping of information out of the public domain or the effective determination of \emph{what} personal information is made available to \emph{whom}~\cite{livingstone2006children}. Nissenbaum's theory of Contextual Integrity (CI) frames privacy as  ``the right to appropriate flow of personal information"~\cite{nissenbaum2004privacy} where what is ``appropriate'' is based on particular contexts and relationships.  CI has been found to be a useful practical framework to interpret people's perceptions of privacy~\cite{barkhuus2012mismeasurement,kummer2016private}.

Managing privacy is becoming increasingly challenging, given the vast (and growing) information asymmetries of the digital age~\cite{acquisti2015privacy}, where data about people are harvested by powerful platforms and vast networks in ways they do not understand nor recognise. Children are particularly susceptible in part due to them having little sense of the risks posed by the accumulation of personal data over time~\cite{Pangrazio:2017:MDM:3097286.3097338}. \comm{The combination of such a lack of understanding, with incentives put in place by apps and platforms to get people to share their data, have yielded the so-called `privacy paradox', a behaviour studied in teenagers~\cite{brandimarte2013misplaced} and adults~\cite{barth2017privacy} alike, in which individuals act in ways contrary to their stated privacy preferences and concerns~\cite{jones2009us,Jia:2015:RLP:2675133.2675287,Pangrazio:2017:MDM:3097286.3097338}.} Recent studies showed that although teenagers from 14 to 18 were typically concerned about being personally identified by their personal data, they failed to perceive the  potential threat of re-identification via the particular fragments they shared, e.g., images or geo-location~\cite{Pangrazio:2017:MDM:3097286.3097338}. 

Whilst there has been extensive research on \comm{teenagers'} perceptions of privacy, relatively little has been done for children aged below 11. They have often been perceived as being too young to understand privacy or exercise digital independence online~\cite{livingstone2018parentsc}.  However, developmental research has shown that children can develop a ``theory of mind" by the age of 4 \cite{colwell2016secret}, which enables children to recognise the differences between concepts in their minds and those in others, and hence to grasp complicated concepts like ``secrecy''~\cite{colwell2016secret} or ``deception''~\cite{chandler1989small}. Contrary to common expectations, children value their privacy because this enables them to enjoy their experience online \cite{Silva:2017:PCT:3091478.3091479, kumar2018cscw}, such as socialising with their friends and families or experimenting with new games.

Recent research with 14 Canadian families showed that children aged 7-11 can align online privacy with real-world scenarios such as ``being left alone" or ``hiding secrets", and draw an analogy to the online environment through expressions like ``keeping things to yourself" or ``not talking to strangers" \cite{zhang2016nosy}. Kumar et al.~\cite{kumar2018cscw} unpacked the perception of online privacy for 26 US young children and examined how they were able to recognise sensitive entities (such as passwords) but struggled with identifying risky actors (strangers) involved in questionable contexts. These studies provide useful insights into children's perception of risks, i.e. what they think of as being risky, and what they struggle to recognise; however, they do not tell us how children make these judgements - or whether children can in fact identify a threat and its origination.

\subsection{Managing Children's Privacy Online}
Tools developed for managing children's privacy on mobile platforms \comm{usually comprise} features that monitor and restrict children's online activities~\cite{wisniewski2017parental,livingstone2018parentsa,ofcom2017}. Children often find these tools overly restrictive and invasive of their personal privacy \cite{Ghosh:2018:SVS:3173574.3173698,wisniewski2017parental}. Co-design studies with children under 12 show that they prefer technologies that facilitate their risk coping skill development, promote communication and interaction with their parents, and emphasise restrictions around monitoring~\cite{mcnally2018co}. However, such tools are still scarce. 

Several recent studies have recognised the importance of supporting children's learning through approaches like interactive storytelling~\cite{Zhang-Kennedy:2016:TIE:2930674.2935984}, game playing~\cite{maqsood2018day} or co-learning with parents~\cite{hashish2014involving}. They have shown the effectiveness of increasing children's awareness of related online safety issues, such as online personal identity or content appropriateness. This demonstrates that a knowledge scaffolding approach to  a child's means of dealing with privacy risk coping could provide a useful addition to safeguarding their cybersafety.

\subsection{Developmental Stages of Privacy Concepts}

Children's ability to recognise privacy risks may be intrinsically limited by their developmental stage. At ages 3 to 5 years, children start to build up friendships; however, at this stage, family interactions are dominant, and many of their online activities are still parent-guided \cite{digital2017}. At this stage, peer pressure is less of an influence~\cite{ann2000developmental}. Children from 6 to 9, \comm{however}, start to learn about the complexity of relationships and feel the need to fit in to peer social groups~\cite{ann2000developmental}. They are also more involved in online activities and enjoy playing games with their friends~\cite{digital2017}. This social interaction with peers makes them more aware of interpersonal privacy risks, but less of other privacy risks \cite{digital2017}. So while they often care deeply about their personal information being shared with their peers, parents, and others online, they remain unaware of other actors, including platforms, app designers, malicious actors, and others operating in digital ecosystems~\cite{livingstone2018childrenc,Pangrazio:2017:MDM:3097286.3097338}. 

Vygotsky's Zone of Proximal Development (ZPD) is a theory that relates the difference between what learners can do independently to what can be achieved by through guidance by a skilled partner~\cite{chaiklin2003zone}. It has been applied to assess the effectiveness of teaching and learning~\cite{miller2010theories}, to identify key barriers in learning by understanding knowledge gaps~\cite{daniels2005introduction}, and in the design of better intelligent tutoring systems (ITS)~\cite{quintana2004scaffolding}. \comm{Inspired by ZPD, we focus on identifying children's misconceptions of privacy risks and highlighting areas of knowledge that could be developed through future active scaffolding. }

\section{Study Design}

Given our focus on understanding children's ability to recognise privacy-related contexts by examining how they describe them, we chose the focus group method to elicit children's responses to a collection of hypothetical scenarios that reflect different types of explicit and implicit threats to children's online personal data privacy. 
\subsection{Focus Groups with Children}

The role of children in research is increasingly recognised, as children are key stakeholders in modern digital technologies~\cite{druin1999role}. Focus groups put children at the centre of the research, \comm{and encourage the sharing of their perspectives and experiences \cite{gibson2007conducting, morgan2002hearing, greene2005researching}}. They are often used to facilitate group dialogues around the topic of interest and result in findings that cannot be obtained through individual interviews~\cite{stewart2014focus}. Focus groups have been successfully used to study children's experiences in different fields, such as health sciences (e.g. studying children's experience of living with asthma~\cite{morgan2002hearing}), in sociology (e.g. studying children's experiences with bullying~\cite{mishna2009ongoing,guerra2011understanding}), or in understanding children's experiences with technology~\cite{ey2011exploring,fane2018exploring,dunn2018s}.

These previous studies have demonstrated that using focus groups with children can reduce the influence of adults, and encourage children to keep each other on track and truthful. However, conducting focus groups can be challenging, given the power dynamics amongst the children and the different forms of expression preferred by different children. Based on the experiences of several previous focus groups with children under 11~\cite{morgan2002hearing,ey2011exploring}, we took the following decisions in our study design: 1) keeping children from the same age group together; 2) using role playing to relax children and balance the power between the adult and children; and 3) providing options for children to express themselves using other methods like pen and paper.

\subsection{Scenarios}
\begin{figure*}[h!]
\centering
\includegraphics[width=0.8\linewidth]{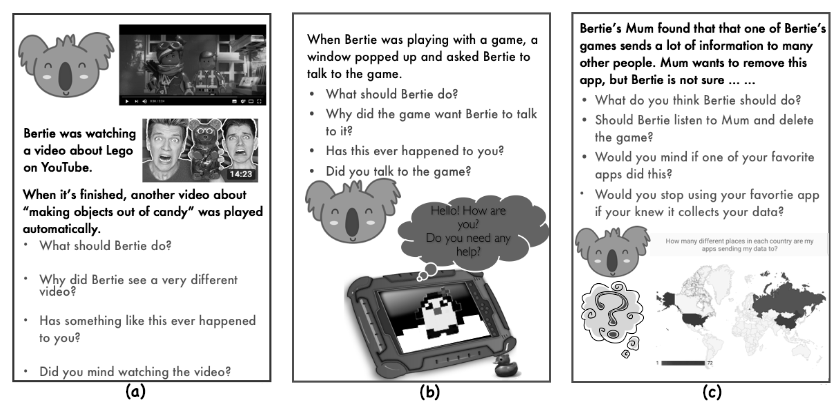}
\caption{Three hypothetical scenarios for focus group discussion (original story cards were in colour): (a) shows a video promotion scenario; (b) illustrates an in-app pop-up requesting children's inputs; and (c) depicts how children's personal data could be tracked and sent to third parties without their knowledge or explicit consent. }\label{fig:stories}
\end{figure*}

In our focus group discussions, we chose to walk through a series of hypothetical scenarios with participant children, about a cartoon character named Bertie, an 8-year-old koala bear who likes playing with tablets, but is not always certain how to cope with unusual events taking place during his use of the mobile apps (see Figure~\ref{fig:stories}). Hypothetical scenarios are effective ways of collecting children's perception of online security and privacy~\cite{kumar2018cscw}, by making them feel they are not being judged~\cite{yao2017folk,wash2010foLk}. 

\comm{Our scenarios were carefully designed to contrast explicit versus implicit data collection in a familiar versus unfamiliar technology context.} Previous research has shown that children under 11 particularly struggle to understand risks posed by technologies or comprehend the context of being online~\cite{zhang2016nosy,kumar2018cscw}. They have explored how children responded to \textit{explicit} data requests such as in-app pop-ups; while our study also looks into children's awareness and perception of \textit{implicit} data access through third-party tracking, which leads to personalised online promotions. 

Story 1 --- Implicit video promotions are widely found in applications like social video sharing platforms, which can be based on personal viewing history or a viewers' interests. Online promotions on such platforms are a significant means by which children discover games or video channels, material that is not always appropriate for their age or developmental needs~\cite{livingstone2017children}. \comm{Therefore, story 1 aimed to assess how much children are aware of the video promotion behaviours of online platforms, some of which could be based on children's online activities, including the videos they have watched or the games they enjoy playing.} 

Story 2 --- In-app pop-ups can explicitly prompt children for personal information (such as names, age or voices) before they can continue with the game. \comm{Story 2 aimed to assess children's awareness of \textit{explicit} stranger danger and in-app game promotions, which can be personalised based on their online data.} 

Story 3 --- The large number of applications (`apps') that can be downloaded for free are a major way by which children interact with these devices. Currently these `free' apps are largely supported by monetisation of user's personal information~\cite{acquisti2016economics,kummer2016private}. A large amount of personal information and online behaviour may be collected from children's apps and shared with third party online marketing and advertising entities~\cite{reyes2017our}. This scenario is designed to examine how \comm{children perceive and feel about these risks}. 

\comm{Scenario 1 and 3 are closely related to children's ability to comprehend the impacts of data-driven decision-making, i.e. their `big data' literacy~\cite{d2015approaches}. Previous research has shown that children aged 8-16 were capable of critiquing analytics applied to the data about them~\cite{hautea2017youth} and recognise issues related to algorithmic accuracy and fairness. Our study complements previous research by looking into how much children aged 6-10 are aware of the context in which their personal data may be collected, transmitted to other online actors, and then used to drive algorithm-based decision-makings.}

\section{Study Method}
Participants were recruited from both local primary schools and a public forum for advertising local family events. Recruitment started in May 2018 after obtaining ethics approval, and 12 studies were carried out between June and August 2018. Half of the studies took place on school premises during school term time, and the other half in our university premises during the summer holiday period. At school, a school teacher or teaching assistant was present in the room, at the university, parents left the children in the room and waited outside until the study was completed. The majority of the participants were recruited through schools and school newsletters. Each study was facilitated by one researcher along with two note-takers. 

\subsection{Study Process}
Each focus group study contained four parts, including a warm-up and introduction session, a sharing of favourite apps, a walk-through of the hypothetical risk scenarios, and finally an open-ended session about issues not so far discussed. The whole study was planned to last around 1 hour and each part of the study was designed to be fun and framed as a game. We avoided using words like `dangerous' or `suspicious' in our scenarios. 

The warm-up session included a game of ``throwing a ball"~\cite{morgan2002hearing} and invited everyone in the room to share their favourite colour and food with the others. The session normally lasted 10 minutes. It helped participants, particularly young children, to relax. Then we used an iPad to ask children to show us their favourite apps, or search for them in the app store if they could not find them on our device. Children were then asked to comment on why they liked the app. This gave us insight about what children in our study enjoyed doing online and their positive experiences.

During the session discussing the three hypothetical scenarios, we let everyone choose a soft toy in the room and role played a character each, such as Bertie's sibling, parent or teacher, in order to help participants (particularly quiet children) feel they are not being judged~\cite{morgan2002hearing}. In each hypothetical scenario, children were first given some time to read through the printed story cards (as illustrated in Figure~\ref{fig:stories}). If they had trouble, one of the note-takers would provide help and read along with them\footnote{Reading literacy is widely promoted in the U.K. for children aged 4 years and above. The story cards were written in as simple a language as possible and were tested with children aged 6yo.}. While children were first asked to respond to the predefined questions, like \textit{what will you do} or \textit{what do you think Bertie should do}, our facilitator followed up any responses that required further clarifications (such as what do you mean by ``hacking''). This session also encouraged children to share their personal experience related to the scenarios, by asking questions like \textit{whether this has happened to them} and \textit{what you did}. This enabled us to listen to children's descriptions of their experiences not covered in the scenarios. 

Audio and video recordings were taken during the studies; two note-takers provided additional details about key interactions they observed. Video recordings were played back during the transcription and data analysis phases to highlight any notable interaction patterns between participants.

\subsection{Participant Information}
We had 29 participant children, including 14 boys and 15 girls, with an average age of 8.5 (range = 6-10, s.d. = 1.4). Details about participants can be found in Table 1. In total we had 12 focus group and the group size varied between 2 and 4~\cite{freeman2008researching}, with an average group size of 2.4. 

Children in our study were generally more attracted to `fun' games that they found exciting or could learn new things from. They were also influenced by friends and families on the choice of apps (`so that I can play with friends'). 25/29 children owned their tablets or phones, while the rest used their parents' devices or shared with their siblings. Online promotions (e.g in-app promotions, app store adverts, YouTube videos) had influenced many children's choice of apps; 20/29 children reported that they have seen promotions in their games or videos, and 12 children reported that they found their favourites or installed new games through these means. 

\begin{table}[h!]
\centering
\begin{tabular}{c|c|c|c}
\rowcolor{Gray}
\textbf{Age}   & \textbf{\#Boys} & \textbf{\#Girls} &\textbf{\#Total}\\ \hline
6-yo  & 4      & 0    &4   \\ 
\rowcolor{Gray}
7-yo  & 1      & 0    &1   \\ 
8-yo  & 3      & 3    &6  \\ 
\rowcolor{Gray}
9-yo  & 3      & 4    &7  \\ 
10-yo & 3      & 8    &11     
\end{tabular}
\caption{Summary of participants' ages and genders }
\vspace{-11mm}
\end{table}

\section{Data Analysis Method}
We transcribed the interviews and analysed the data in two iterations. In the first iteration, we used a thematic analysis method to develop codes and themes related to risks talked about by the children and how they coped with them. 

\comm{The thematic coding process started by dividing the transcriptions into two (roughly) equal-sized sets. The first three authors independently analysed the first set of transcriptions to derive an initial set of codes. Then they met to consolidate and reconcile codes into a common codebook. These codes were then applied to the second (yet unseen) set of transcriptions by the same set of researchers, and Fleiss' kappa~\cite{fleiss1971measuring} (0.83) was computed to assess inter-coder reliability.}

\comm{In the second iteration, we used the CI framework, particularly its four-parameter model, to examine the risks identified by the children (either from our hypothetical scenarios or their own experiences with technologies) and how they were described by the children. The four parameters from the CI framework include the following:
\begin{itemize}
    \item \textit{Attributes}: the types of information being transmitted, such as users' data, personal information, etc.
    \item \textit{Contexts}: the situation or scenario to which the social norms may be applied. 
    \item \textit{Actors}: the entities involved in the information transmission, which can be the subject, sender or recipient of the information. 
     \item \textit{Transmission principles}: the way information is transmitted from the sender of information to the recipients.
 \end{itemize}}

\comm{This four-parameter model helped us to calibrate whether children recognised the exact actors, attributes, context and information transmissions involved in each risk context, led us to delineate three categories of risk recognition by children, namely \textit{accurate, vague and missed} risk recognition.}

\section{Results}
\comm{We present our results by first outlining children's use of language under each category of risk recognition (i.e. recognised, vaguely or missed). We then present an overview of children's responses to the three hypothetical scenarios, before providing in-depth analysis regarding children's risk coping strategies when risks were recognised, vaguely understood or missed. In this way, we demonstrate an understanding of children's current knowledge gaps in describing and coping with risks.}

\subsection{Children's Use of Language Under Each Category of Risk Recognition}

\begin{table*}[h!]
\resizebox{\textwidth}{!}
{
\begin{tabular}{llll}
\rowcolor{Gray}
Context & Risks & Words & Children's explanation of the meanings    \\ \hline
        & Inappropriate content          & \textit{weird things}                             & \textit{things for adult, not my age} \\
        & Threats from strangers         & \textit{strangers,}                               & \textit{random people, or I don't know} \\
\multirow{-3}{*}{Risks recognised}       & Person information over-sharing & \textit{personal information} & \textit{everything about me, my stuff, personal photos} \\ \hline
        & Online promotions              & \textit{Channel people, app developers}  & \textit{I don't know} \\
\multirow{-2}{*}{Risks vaguely understood} &                                & \textit{get more subscribers} & \textit{more money and more famous} \\
& Pop-ups   & \textit{hacking} & \textit{steal from house, take your account, steal your data}                                 \\
& Data tracking                  & \textit{hacking} & \textit{track your personal information}                                                      \\
    &   & \textit{tracking} & \textit{try and find you, find your location, know more about what's happening in this country} \\ \hline
No risks specified  && \textit{scary, angry, upset, annoying, surprised} &   \\\hline                  
\end{tabular}
}\caption{Example descriptions of risks felt, experienced, recognised or interpreted by children.}
\label{tab:words}
\vspace{-3mm}
\end{table*}

We observed that some terms were repeatedly used by children across different situations. Moreover, as shown by the examples in Table~\ref{tab:words}, children were able to describe risks accurately when they could recognise the actual risks. However, when they had only a vague understanding of the risks, they struggled to describe things consistently or to provide a good explanation of what they meant. 

\begin{flushleft}
\textit{Good risk recognition and accurate description}\\
\end{flushleft}
When children recognised the actual risks, such as inappropriate content or over-sharing of personal information, they could describe them quite accurately, for example, using terms like \textit{`personal information'} or \textit{`private information'} to refer to the type of information they treated as sensitive; or words like \textit{`things for adults'} or \textit{`not my age'} to refer to content that was inappropriate to them.\\
\vspace{-3mm}
\begin{flushleft}
\comm{\textit{Vague risk recognition and inconsistent descriptions}}
\end{flushleft}
\comm{When children only made a partial sense of certain types of risks, their use of terms can be inconsistent.}
For example, although some children were able to describe the scenario of new videos being presented to Bertie using words like `people trying to make money' or `them trying to make you watch more', they struggled to explain who these `people' are and how information might be transmitted to these `[YouTube] channel people' in this context. Another example is the term `hacking', which has been used by children across different focus groups, but in fact with very different meanings. For example, when trying to explain why a new video was shown following a previous video, children used `hacking' to mean `someone stole my data', `take your account', or `steal from house'. `Hacking' has been used by the same group of children to make sense of other scenarios, including in-app pop-ups or data tracking.\\

\vspace{-3mm}
\begin{flushleft}
\textit{Feelings and experiences, but no recognition of risks}\\
\end{flushleft}
When children could not describe the exact risks they encountered, they would refer to a feeling of `scary' or `annoying' to talk about times when they felt a need to take an action. Children also referred to things they `had experienced before' when they could not describe why something made them take actions. This indicates that although sometimes children probably did not \comm{fully comprehend} their previous experiences, the consequences provided a signpost for them to stop~\cite{Jia:2015:RLP:2675133.2675287}.

\subsection{\comm{Children's Responses to the Hypothetical Scenarios}}

\begin{table}[ht!]
\resizebox{\columnwidth}{!}{
\begin{tabular}{c|p{0.9in}|p{0.8in}|p{1in}}
\rowcolor{Gray}
\textbf{Context} & \textbf{Bertie should?} & \textbf{You would?} & \textbf{Why?} \\ \hline

Story 1 & Play-and-see, or tell parents & Play-and-see, or continue (because it is fun) & \textit{autoplay, making it easier (to watch), based on past history, people making money}\\
\rowcolor{Gray}
Story 2 & Stop, delete, or tell parents  & Shut off or delete & \textit{hack you, trick you, steal the voice, access personal information} \\

Story 3 & Stop, tell parents, or do some research & Stop & \textit{my information, depending the companies} \\
\end{tabular}}
\caption{Children's responses to hypothetical scenarios --- what to do and why}
\label{tab:scenarios}
\vspace{-8mm}
\end{table}

Table~\ref{tab:scenarios} summarises how children responded to our three hypothetical scenarios:
\begin{itemize}
    \item In the video-related scenario of story 1, \comm{which aimed to assess children's awareness of (personalised) game promotions,} children largely \textit{missed} these risks and associated this context with `autoplay' or computers trying to `save your time'; only three groups associated this context with YouTubers trying to `make money'. 9 out of 12 groups suggested that they would `play and see', and half of the groups suggested that Bertie should tell his parents. However, when asked what they would do in this scenario, all but one of the groups would play the video and make a judgment by themselves, instead of seeking advice from their parents.
    \item \comm{Story 2 aimed to assess children's awareness of stranger danger and in-app game promotions, which again can be personalised}. Apart from one group, all children agreed that \textit{pop-ups should be treated with caution}, whilst they made different interpretations of pop-ups, which could be `hacking' (i.e. `stealing their information or voice data') or `tricking them to buy things'. Even though the perception of risks differed, all children suggested that both Bertie and they should go and tell someone about this, or immediately delete or stop using the app.
    \item \comm{For story 3, which focused on the tracking behaviour of apps, an aspect still largely unfamiliar to adults and children alike}, `telling parents' and `discussing with parents' were suggested by 10/12 groups as the coping strategies, even though children didn't fully understand who might access this tracked data. A few older children (9-10 yo) also suggested that they would do some further research with their parents to figure out what kind of information might be collected or who the companies might be that were receiving this information. All children said that Bertie should stop using the app and they would stop using their favourite apps too.
\end{itemize}

\begin{table*}[h!]
\small
\begin{tabular}{lll}
\rowcolor{Gray} 
Context                                    & Descriptions and examples                                                                                                                    & Children's risk coping strategies                                                        \\ \hline
Risks recognised                           & \begin{tabular}[c]{@{}l@{}}inappropriate content\\ \\ request for sensitive personal information \\ \\ approaching by strangers\end{tabular} & \textit{\begin{tabular}[c]{@{}l@{}}ask for help\\ stop\\ avoid oversharing\end{tabular}} \\ \hline

    & Online promotions == recommendations & \textit{it's ok, let's play}                                                             \\ \cline{2-3}
  & Tracking == hacking & \textit{\begin{tabular}[c]{@{}l@{}}stop\\ ask for help\end{tabular}}                     \\ \cline{2-3}
\multirow{-3}{*}{Risks vaguely understood}                                           & Pop-ups == hacking                                                                                                                           & \textit{\begin{tabular}[c]{@{}l@{}}stop\\ ask for help\end{tabular}}                     
 \\ \hline 
                                           & New videos == auto play                                                                                                                      & \textit{it's ok, let's play}                                                             \\ \cline{2-3}
                                           & Familiar YouTuber/games == Ok                                                                                                                & \textit{it's ok, let's play}                                                             \\ \cline{2-3}
\multirow{-3}{*}{Risks missed}             & I/My friends played it before == Ok                                                                                                          & \textit{it's ok, let's play}    \\ \hline                                                        
\end{tabular}\label{tab:coping}\caption{Children's Risk Coping Strategies Depending on Their Ability to Recognise Risk Contexts}
\end{table*}

\subsection{Risk Coping When Risks Were Accurately Recognised}
\comm{Children demonstrated a strong consciousness of their online identity and the importance of avoiding sharing their real identity or over-sharing their personal information. In these cases,} children applied various effective strategies to protect their sensitive personal information. For example, a 10-yo girl demonstrated her knowledge of using an obfuscation strategy to protect her real identity online:
\begin{quote}
`I make up a name cause I don't want people know my name' -- C6, a 10-yo girl
\end{quote}

Others mentioned techniques they would use to verify the identity of anyone who tried to contact them online:
\begin{quote}
`If I know who they are and they told me their names and accounts, and if they ask me in person to friend them, then I would friend them. If I don't know who they are and haven't seen them in real life then I wouldn't accept them' -- C14, a 9-yo girl
\end{quote}

\comm{These risk coping strategies are effectively applied by the children when they face explicit inappropriate content or an explicit request for their personal information from platforms or apps}. However, our analysis showed that children may struggle to fully understand risks in other contexts. In the following sections, we unpack how children responded to situations when they struggled to recognise or understand the risks fully, summarised in Table 4.

\subsection{Risk Coping When Risks Were Vaguely Understood}
\comm{In contrast to explicit privacy risks, children struggled to associated online promotions with losing control of personal information. The effectiveness of their risk coping may vary --- they may be cautious even though not fully understanding risks, or open to try-it-and-see strategies.}

\begin{flushleft}
\textit{Open to Recommendations}
\end{flushleft}
Several children discussed their interpretation of how YouTubers may try to `persuade' them to watch their videos in order to gain `money' or `more subscribers'. For example, a group of 8-yos shared their knowledge about how the number of subscribers to an online video could be related to the reward to the video publisher.
\begin{quote}
`C12: eurgh, they get money and they get more famous\\
C10: if you watch their video a lot, then they get a lot of thumbs ups of their videos \\
C13: they have lots of subscribers'
\end{quote}

However, none of these children expressed resistance to these video promotions. They treated it as if this is how the Internet works -- `if they reach the max subscribers, then they get the money' (C12). Their primary decision making was still based on whether the content was interesting or whether it was from their favourite video providers. \\ 
\vspace{-3mm}
\begin{flushleft}
\comm{\textit{Play-and-See When Not Fully Recognising Recommendations}}
\end{flushleft}
Only a few children (3/12 focus groups) recognised \comm{how new videos might be related to personalised recommendations}. One group of 10-yo children used the word `recommendation' to describe video promotions on YouTube. However, they struggled to understand \textit{who} was performing these recommendations, and as a result, they were less sure about the consequences for their privacy.
\begin{quote}
`C2: because it recommended this. \\
C1: maybe he watched other videos like that. and then this one popped up \\
R: how did you know about that? \\
C1: not sure'
\end{quote}

As a result, the children usually applied the play-and-see strategy to assess the content or apps they came across, without any more complex reasoning.\\

\begin{flushleft}
\comm{\textit{Stop, Even Not Fully Recognising What Hacking Involves}}
\end{flushleft}

Children associated a diverse range of scenarios with the word `hacking'. Sometimes it was used deliberately to refer to actions (or intentions) by app and platform designers, rather than activities of computer criminals. For example, the following description from a 10-yo girl referred to her experience with location tracking by Snapchat. She understood risks associated with location tracking, even though she used the word `hacking' to describe tracking behaviour of the app:
\begin{quote}
`Yeah, that's why I put on ghost mode, so they can't find your location. So yeah, they try to hack your tablet and they can get all the games you like to play in, all the personal information, like what school you go to' -- C5, a 10-yo girl
\end{quote}

Other times, children might use the word to refer to being personally coerced or made to take an action by an app or service.  For example, C24 (a 7-yo girl) tried to explain why she thought Bertie should not watch the new video by saying that
\begin{quote}
`It is a bit random and just pops up ... someone might be trying to hack you or something'. -- C24, a 7-yo girl
\end{quote}
In such a situation, the concept was `hacking' was used to provide a possible explanation as to \emph{why} or \emph{by whom} they were being made to take particular actions--which they struggled to fully understand. 

This was particularly common among younger children under 8 (11/29), who could not explain what they meant by `hacking' when this word was used in their interviews. In these situations, because children recognised that `hacking' is a bad thing, they would take effective action to either stop or ask for help from parents, even though their actual recognition of the risks were vague or misinformed. 

\subsection{Risk Coping When Risks Were Missed}
Children could miss risks due to a lack of knowledge, or due to their past experiences, which did not lead to any direct consequences related to the risks. For those children who interacted with certain technologies and experienced no implications before, they would tend to be more (over-) confident with technologies.

\begin{flushleft}
\textit{Associating New Videos with `Autoplay'}\\
\end{flushleft}
Children from 7 focus groups treated online video promotions as part of an `autoplay' function of the platform, without questioning how the new content might be presented to them. 12/29 children demonstrated trust in the content provided by their familiar YouTubers. For example, C4 (a 10-yo girl) mentioned that `because it's one of my favourite YouTubers. So I was ok with it'. 

As a result, children reported having been exposed to unexpected content and online baiting.  These same children reported that they often saw upsetting content online (e.g. `sometimes in autoplay, it comes up with these really freaky ones like pictures of dead people ' --- C5, a 10-yo girl). 12 (out of 29) children in our study reported how their favourite games were discovered through promotions in the videos they watched (`A friend and I were watching YouTube and we saw people playing this game.' -- C14, a 8-yo girl). 
\begin{flushleft}
\textit{Familiarity Overwrites Rules}\\
\end{flushleft}
Our data also shows that familiarity could give children a fake sense of safety. For example, C3 (a 10-yo girl) mentioned that `I don't think YouTube and stuff like that could collect much', and another two 10-yo girls discussed how their experiences with a known app had a strong influence on their judgement upon whether an app could pose threats to them or not.

\begin{quote}
`C8: It depends on what game it is. If it's like the talking Tom, then that's fine because you're safe.\\
C9: Yes, cause it doesn't record you and keeps it.\\
C8:Because I play it all the time and nothing has happened to me and I'm always talking to it. I haven't had any problem with my ipad\\
c9: Yes me too, I've been playing with it for one to two years and haven't had any problem.'
\end{quote}

\begin{flushleft}
\textit{Experience-Centric Decision Making}\\
\end{flushleft}

We also observed that a child's or their peers' personal experiences had a strong  influence upon their decision making, even though they didn't always understand what may pose threats to them. The following example illustrates how children in our study demonstrated that they used experiences as a heuristics for their decision making, as described by C27, a 10-yo girl, `If my friends already play it I wouldn't bother checking it cause they know it's safe. If it's one I didn't know about then I wouldn't use the app'. 

\section{Discussions}
\subsection{Key Findings and Contributions}
Our results reinforce existing findings that children under 11 struggle to fully understand online privacy risks~\cite{byron2010we,ey2011exploring,digital2017,zhang2016nosy,kumar2018cscw}, especially those associated with implicit personal data collection and use, through mechanisms such as data tracking or in-app recommendations. 
However, our results also demonstrate that children cared about, and were sensitive to, who might access their sensitive information (e.g. real names, age, location etc), and applied a range of techniques to safeguard this space, such as by verifying identities through face-to-face interactions or avoiding using real names as usernames. Children felt `annoyed', `surprised' or `angry' when they felt coerced, or felt not in control. Our results also reinforced that, like teenagers and younger children, our participants valued the positive experiences of being online and keeping in touch with their friends, and their personal space online~\cite{lwin2008protecting}. 

\subsection{Implications for Designing for Children}
It is clear that current digital technologies are often not designed with children's best interests in mind. For example, social media platforms widely used by children (such as Instagram or Twitter) keep their profile pages as public by default\cite{pri}, and games from the `family' genre from the leading mobile app markets are associated with more third-party trackers than games designed for adults~\cite{Binns2018MobileEcosystem,reyes2018won}. For younger children, smart toys and connected baby monitors can continuously stream video and audio information to data centres in ways that are completely opaque to children and/or parents~\cite{mcreynolds2017toys}.

Although the need to provide a better design for children is recognised by the recent development of regulations (such as the General Data Protection Regulation (GDPR) \cite{chigdpr}) and proposals for age-appropriate design for children \cite{ico}, children are only occasionally consulted in these efforts \cite{usapp_2017}. Designers, as well as policymakers, should recognise children's interests as well as the need to involve children~\cite{mcnally2018co,kumar2018co} in the process of design and policymaking.

\subsection{Implications for Privacy Tools Supporting Parents and Children}

Our findings confirmed children's ability to recognise certain privacy risks. But we need to expand children's understanding of implicit data collection risks. Current tools for safeguarding children online mainly focus on enabling parents to take control or monitor children's online activities~\cite{wisniewski2017parental,kumar2018cscw}. Related research with parents and their teenagers have shown that current tools often work against parents' and children's values of privacy, and they would prefer tools to facilitate parental mediation of children's use of technologies rather than simply providing surveillance capabilities~\cite{wisniewski2017parental,wisniewski2018privacy}. \comm{Parental involvement leading to improved learning outcomes of children is extensively supported by existing literature~\cite{cheung2011parents,whalley2017involving}; however,} parents feel they are poorly supported in dealing with challenges related to facilitating their children's use of digital technologies~\cite{livingstone2018parentsa}. Most of the time they rely on self-guided online searches, rather than being informed by systematic, comprehensive and reliable resources~\cite{livingstone2018parentsa}. \comm{Future tool development should consider both scaffolding children's knowledge acquisition and facilitating the active involvement of the parents.}

Several studies have looked into how to facilitate children's learning of online privacy and safety through a co-learning process between parents and children~\cite{zhang2017cyberheroes,hashish2014involving}, and demonstrated their effectiveness for increasing children's risk recognition and coping. However, the difficulty of recognising data tracking risks is not unique to children, suggesting the need for better tools to help provide some transparency to adults and children alike~\cite{acquisti2015privacy, vankleek2018chi}. \comm{Greater transparency could also support children's ability to make sense of the personalisation and recommendation behaviours of online platforms that dominate our information consumption online~\cite{d2015approaches}, by providing information about why or by whom particular content was recommended.} However, given children's development stages, especially those under~11, they probably have less ability to fully recognise the implications of these risks. Therefore, tools designers should not only help parents mediate children's understanding of online risks but also help them to develop \comm{``big data'' literacy~\cite{hautea2017youth} to start to understand how information derived from online activities are captured, retained and repurposed by various entities, as well as the potential risks such retention, processing, and use carry.}

\subsection{Reflection on Methods}

Focus groups have been instrumental in our understanding of children's perception of risks and their way of describing these risks. This approach has provided an open setting to incentivise children's sharing of their experiences, and particularly those of their peers, like friends or siblings. In comparison to previous studies~\cite{kumar2018cscw,zhang2016nosy}, in which semi-structured interviews with parents and children were the main method used, our focus group studies with children inspired more discussions about children's and their peers' experiences with online privacy risks. These have been a principal means for us to have a deeper understanding of how children's risk coping strategies may vary under different contexts.

\comm{The CI framework provides a useful philosophical framework for understanding the social and ethical aspects of privacy. However, translating it into a technical implementation is not straightforward~\cite{benthall2017contextual}. Most of the previous work has used \textit{context} as a synonym for a scenario or situation, to understand users' privacy preferences and expectations~\cite{wijesekera2015android,zhang2013no}. Others have used the concept to design social platforms that can adapt to different social norms in different spheres (like families, friends, colleagues etc)~\cite{criado2015implicit,tierney2014realizing}.}

\comm{Kumar et al.~\cite{kumar2018co} have applied the normative aspect of the CI framework to assessing how well children can recognise the \textit{actors, information entities and information transformations} involved in a context. This work extends and builds on their findings:  by examining children’s privacy mental models, we focused on the terms used by the children for describing privacy contexts and related factors (e.g. the information that is sensitive to them or the organisations that are regarded as threatening). We look for a deeper understanding of both the types of risks that children struggled to recognise (such as third-party tracking) and to describe these accurately.} This has been an effective process to understand children’s current knowledge gaps and the key information that is needed to facilitate their future development of privacy knowledge.

\section{Limitations and Future Work}

We acknowledge that it is difficult to generalise our findings given the sample size and study population. First, schools and parents who chose to participate in the study may already be more interested in the topic than the average population. This may have impacted the \emph{a priori}  awareness of online privacy risks our participants had. Secondly, whilst we did not collect information about participants' family income, the families' areas of residences were centred in an affluent area near a university. This may have resulted in our findings reflecting a greater privacy literacy than the general population, not only due to familial influences, but also due to local schools. Our primary goal, however, was not to measure the degree of literacy or concern, but the gaps of greatest concern and potential for new tools to help. \comm{While we did not explicitly look at age-specific differences, informally we noticed possible differences we wish to examine further in future work; e.g. children older than eight seemingly demonstrating a richer vocabulary to describe risks.} Finally, when the focus group studies were carried out at schools, a teaching assistant or a class teacher was often present in the room, for safeguarding reasons. Children may have moderated their responses in these settings.  

Future work also aims to explore how we may design approaches that could enable children to expand their knowledge about online privacy risks in two ways: \comm{new knowledge scaffolding based upon children's actual understanding, and addressing the critical knowledge gap about online promotions and data tracking.} \comm{Our study emphasises the importance of scaffolding children’s understanding of risks and privacy strategies.} Given children's role in these digital technologies, we intend to use co-design workshops~\cite{kumar2018co,mcnally2018co}, involving children, parents, teachers, and designers of educational games, to explore approaches to scaffolding children's privacy knowledge.

\section{Conclusion}
Inspired by the ZPD theory and the Contextual Integrity framework, our work examined children's current knowledge about privacy risks online and how children may or may not have fully recognised online privacy risks when adopting particular strategies. Our results showed that children's ability to fully recognise privacy risks has a direct impact on their ability to consistently describe and manage these risks: when they only vaguely recognised the risks, they would try to make sense out of them using their knowledge or experiences, but would not always take effective action. Expanding our understanding of children's perceptions of risks thus advances the goal of facilitating a child's ability to cope with risks from a young age, scaffolding this through a knowledge acquisition, rather than a restrictive approach. We hope that our findings will support both designers of new  privacy tools for children, as well as those of educational material seeking to address gaps in their understanding of risks and data use online.

Providing better privacy-by-design guidelines for protecting children is essential both to influence, and meet the aspirational goals of data protection (DP) initiatives being set forth around the globe. Although the GDPR in the EU protects the use of children's data, the Children's Online Privacy Protection Act (COPPA) in the US has yet to provide an explicit regulation of third-party tracking. This study has highlighted some potential avenues by which future tools might, through greater data literacy, lay the foundation for having children understand, and start to exercise, the rights such DP regulation grant them. 

\section{Acknowledgments}
\comm{We thank participant schools and children  for their time, and the anonymous reviewers for their feedback. This work was supported by the EPSRC Impact Acceleration Account KOALA (EP/R511742/1).}

\bibliographystyle{ACM-Reference-Format}

\end{document}